\documentclass[reprint, amsmath, amssymb, superscriptaddress, aps,prl]{revtex4-2} 

\pdfoutput=1

\usepackage{graphicx}
\usepackage{dcolumn}
\usepackage{bm}
\usepackage{hyperref}

\def\t{\mathrm}

\begin{document}

\preprint{APS/123-QED}

\title{Learning Efficient Navigation in Vortical Flow Fields}

\author{Peter Gunnarson}
\affiliation{Graduate Aerospace Laboratories, California Institute of Technology, 1200 E California Blvd, Pasadena, California 91125, USA}

\author{Ioannis Mandralis}
\affiliation{Computational Science and Engineering Laboratory, ETH Zurich, 8093 Zurich, Switzerland}
\author{Guido Novati}
\affiliation{Computational Science and Engineering Laboratory, ETH Zurich, 8093 Zurich, Switzerland}
\author{Petros Koumoutsakos}
\affiliation{Computational Science and Engineering Laboratory, ETH Zurich, 8093 Zurich, Switzerland}
\affiliation{John A. Paulson School of Engineering and Applied Sciences, Harvard University, 150 Western Ave, Boston, MA 02134, USA}
\author{John O. Dabiri}
\affiliation{Graduate Aerospace Laboratories, California Institute of Technology, 1200 E California Blvd, Pasadena, California 91125, USA}
\affiliation{Mechanical and Civil Engineering, California Institute of Technology, 1200 E California Blvd, Pasadena, CA 91125, USA}

\date{\today}

\begin{abstract}
Efficient point-to-point navigation in the presence of a background flow field is important for robotic applications such as ocean surveying. In such applications, robots may only have knowledge of their immediate surroundings or be faced with time-varying currents, which limits the use of optimal control techniques for planning trajectories. Here, we apply a novel Reinforcement Learning algorithm to discover time-efficient navigation policies to steer a fixed-speed swimmer through an unsteady two-dimensional flow field. The algorithm entails inputting environmental cues into a deep neural network that determines  the swimmer's actions, and deploying Remember and Forget Experience replay. We find that the resulting swimmers successfully exploit the background flow to reach the target, but that this success depends on the type of sensed environmental cue. Surprisingly, a velocity sensing approach outperformed a bio-mimetic vorticity sensing approach by nearly two-fold in success rate. Equipped with local velocity measurements, the reinforcement learning algorithm achieved near 100\% success in reaching the target locations while approaching the time-efficiency of paths found by a global optimal control planner. 
\end{abstract}

\maketitle
\emph{Introduction}.\textemdash Navigation in the presence of a background unsteady flow field is an important task in a wide range of robotic applications, including ocean surveying \cite{weizhong_zhang_optimal_2008}, monitoring of deep-sea animal communities \cite{kuhnz_benthic_2020}, drone-based inspection and delivery in windy conditions \cite{guerrero_uav_2013}, and weather balloon station keeping \cite{bellemare_autonomous_2020}. In such applications, robots must contend with unsteady fluid flows such as wind gusts or ocean currents in order to survey specific locations and return useful measurements, often autonomously. Ideally, robots would exploit these background currents to propel themselves to their destinations more quickly or with lower energy expenditure. 

If the entire background flow field is known in advance, numerous algorithms exist to accomplish optimal path planning, ranging from the classical Zermelo's equation from optimal control theory \cite{zermelo_uber_1931, techy_optimal_2011} to modern optimization approaches \cite{panda_comprehensive_2020, kularatne_going_2018, guerrero_uav_2013,  weizhong_zhang_optimal_2008, petres_path_2007,lolla_time-optimal_2014}. However, measuring the entire flow field is often be impractical, as ocean and air currents can be difficult to measure and can change unpredictably. Robots themselves can also significantly alter the surrounding flow field, for example when multi-rotors fly near obstacles \cite{shi_neural_2019} or during fish-like swimming \cite{verma_efficient_2018}. Additionally, oceanic and flying robots are increasingly operated autonomously and therefore do not have access to real-time external information about incoming currents and gusts (e.g. \cite{fiorelli_multi-auv_2006,caron_macro-_2008}).

Instead, robots may need to rely on data from on-board sensors to react to the surrounding flow field and navigate effectively. A bio-inspired approach is to navigate using local flow information, for example by sensing the local flow velocity or pressure. Zebrafish appear to use their lateral line to sense the local flow velocity and avoid obstacles by recognizing changes in the local vorticity due to boundary layers \cite{oteiza_novel_2017}. Some seal species can orient themselves and hunt in total darkness by detecting currents with their whiskers \cite{dehnhardt_seal_1998}. Additionally, a numerical study of fish schooling demonstrated how surface pressure gradient and shear stress sensors on a downstream fish can determine the locations of upstream fish, thus enabling energy-efficient schooling behavior \cite{weber_optimal_2020}. 

Reinforcement Learning (RL) offers a promising approach for replicating this feat of navigation from local flow information. In simulated environments, RL has successfully discovered energy-efficient fish swimming \cite{gazzola_reinforcement_2014,jiao_learning_2020} and schooling behavior \cite{verma_efficient_2018}, and a time-efficient navigation policy for repeated quasi-turbulent flow using position information \cite{biferale_zermelos_2019}. In application, RL using local wind velocity estimates outperformed existing methods for energy-efficient weather balloon station keeping \cite{bellemare_autonomous_2020} and for replicating bird soaring \cite{reddy_glider_2018}. Other methods exist for navigating uncertainty in a partially known flow field such as fuzzy logic or adaptive control methods \cite{panda_comprehensive_2020}, however RL can be applied generally to an unknown flow field without requiring human tuning for specific scenarios.

The question remains, however, as to which environmental cues are most useful for navigating through flow fields using RL. A biomimetic approach suggest that sensing the vorticity could be beneficial \cite{oteiza_novel_2017}; however flow velocity, pressure, or quantities derived thereof are also viable candidates for sensing.

In this letter, we find that Deep Reinforcement Learning can indeed discover time-efficient, robust paths through an unsteady, two-dimensional (2D) flow field using only local flow information, where simpler strategies such as swimming towards the target largely fail at the task. We find, however, that the success of the RL approach depends on the type of flow information provided. Surprisingly, a RL swimmer equipped with local velocity measurements dramatically outperforms the bio-mimetic local vorticity approach. These results show that combining RL-based navigation with local flow measurements can be a highly effective method for navigating through unsteady flow, provided the appropriate flow quantities are used as inputs to the algorithm.

\emph{Simulated Navigation Problem}.\textemdash As a testing environment for RL-based navigation, we pose the problem of navigating across an unsteady von K\'{a}rm\'{a}n vortex street obtained by simulating 2D, incompressible flow past a cylinder at a Reynolds number of 400. Other studies have investigated optimal navigation through real ocean flows \cite{weizhong_zhang_optimal_2008}, simulated turbulence \cite{biferale_zermelos_2019}, and simple flows for which there exist exact optimal navigation solutions \cite{kularatne_going_2018}. Here, we investigate the flow past a cylinder to retain greater interpretability of learned navigation strategies while remaining a challenging, unsteady navigation problem. 

\begin{figure}[t!!]
\includegraphics[width=8.64cm]{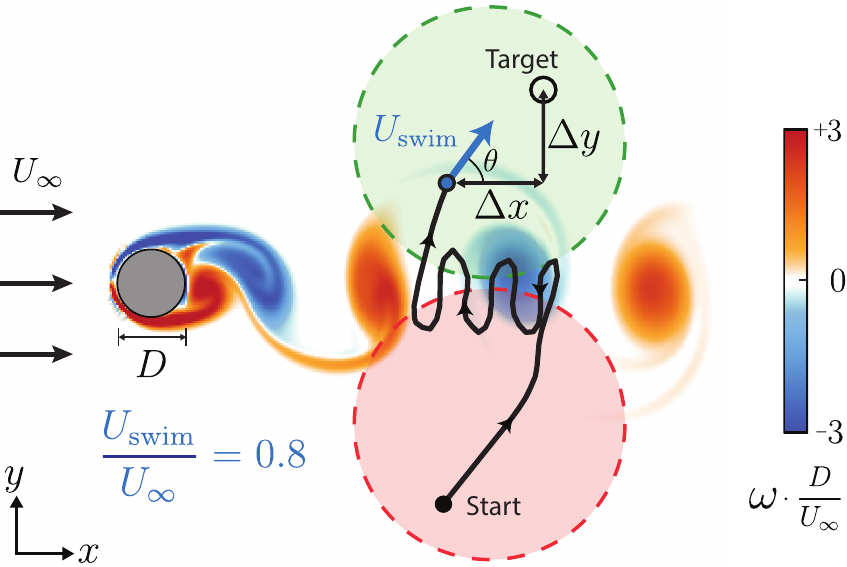}
\caption{Test navigation problem of navigating through unsteady cylinder flow. Swimmers are initialized randomly inside the red disk and are assigned a random target location inside the green disk. These regions of start and target points are $4D$ in diameter, and are located $5D$ downstream and centered $2.05D$ above and below the cylinder. Additionally, each swimmer is initialized at a random time step in the vortex shedding cycle. An episode is successful when a swimmer reaches within a radius of $D/12$ around the target location.}
\label{fig:cylinder_flow}
\end{figure}

The swimmer is tasked with navigating from a starting point on one side of the cylinder wake to within a small radius of a target point on the opposite side of the wake region. For each episode, or attempt to swim to the target, a pair of start and target positions are chosen randomly within disk regions as shown in Figure \ref{fig:cylinder_flow}. Additionally, the swimmer is assigned a random starting time in the vortex shedding cycle. The spatial and temporal randomness prevent the RL algorithm from speciously forming a one-to-one correspondence between the swimmer's relative position and the background flow, which would not reflect real-world navigation scenarios. All swimmers have access to their position relative to the target ($\Delta x$, $\Delta y$) rather than their absolute position to further prevent the swimmer from relying on memorized locations of flow features during training. 

For simplicity and training speed, we consider the swimmer to be a massless point with a position  $\mathbf{X}_n = [x,y]$ which advects with the time-dependent background flow $\mathbf{U}_{\t{flow}} = [u(x,y,t),v(x,y,t)]$. The swimmer can swim with a constant speed $U_{\t{swim}}$ and can directly control its swimming direction $\theta$. These dynamics are discretized with a time step $\Delta t = 0.3 D/U_{\infty}$ using a forward Euler scheme:

\begin{align}
   &\mathbf{X}_{0} = \mathbf{X}_{\t{start}}, \label{eq:start}\\
    &\mathbf{X}_{n+1} = \mathbf{X}_n+\Delta t \left(U_{\t{swim}}\left[\cos{\left(\theta\right)},\sin{\left(\theta\right)}\right] + \mathbf{U}_{\t{flow}}\right). \label{eq:dynamics}
\end{align}

\noindent
It is also possible to apply RL-based navigation with more complex dynamics, including when the swimmer's actions alter the background flow \cite{verma_efficient_2018}. 

We chose a swimming speed of 80\% of the freestream speed $U_{\infty}$ to make the navigation problem challenging, as the swimmer cannot overcome the local flow in some regions of the domain. A slower speed ($U_{\t{swim}}<0.6 U_{\infty}$) makes navigating this flow largely intractable, while a swimming speed greater than the freestream ($U_{\t{swim}}>U_{\infty}$) would allow the swimmer to overcome the background flow and easily reach the target.

\emph{Navigation Using Deep Reinforcement Learning}.\textemdash In Reinforcement Learning, an agent acts according to a policy, which takes in the agent's state $s$ as an input and outputs an action $a$. Through repeated experiences with the surrounding environment, the policy is trained so that the agent's behavior maximizes a cumulative reward. Here, the agent is a swimmer, the action is the swimming direction $\theta$, and we seek to determine how the performance of a learned navigation policy is impacted by the type of flow information contained in the state.

To this end, we first consider a \emph{flow-blind} swimmer as a baseline, which cannot sense the surrounding flow and only has access to its position relative to the target ($s = \{\Delta x, \Delta y\}$). Next, inspired by the vorticity-based navigation strategy of the zebrafish \cite{oteiza_novel_2017}, we consider a \emph{vorticity} swimmer with access to the local vorticity at the current and previous time step in order to sense changes in the local vorticity ($s = \{\Delta x, \Delta y, \omega_{t}, \omega_{t-\Delta t}\}$). We also consider a \emph{velocity} swimmer, which has access to both components of the local background velocity ($s = \{\Delta x, \Delta y, u, v\}$). Other states were also investigated, and are included in supplemental materials.

We employ Deep Reinforcement Learning for this navigation problem, in which the navigation policy is expressed using a deep neural network. Previously, Biferale et al. \cite{biferale_zermelos_2019} employed an actor-critic approach for RL-based navigation of repeated quasi-turbulent flow. The policy was expressed using a basis function architecture, requiring a coarse discretization of both the swimmer's position and swimming direction. Here, a single 128$\times$128 deep neural network is used for the navigation policy, which accepts the swimmers state (i.e. flow information and relative position) and outputs the swimming direction as continuous variables. The network also outputs a Gaussian variance in the swimming direction to allow for exploration during training. The policy network is randomly initialized and then trained through repeated attempts to reach the target using the V-RACER algorithm \cite{novati_remember_2019}. The V-RACER algorithm employs Remember and Forget Experience replay, in which experiences from previous iterations are used to update the swimmer's current policy in a stable and data-efficient manner. Due to the random intialization of the policy, the training process is repeated five times for each swimmer to ensure reproducibility \cite{henderson_deep_2019}. 

At each time step, the swimmer receives a reward according to the reward function $r_n$, which is designed to produce the desired behavior of navigating to the target. We employ a similar reward function as Biferale et al. \cite{biferale_zermelos_2019}:

\begin{equation} \label{eq:reward}
\begin{split}
    r_n = -\Delta t + 10\left[\frac{||\mathbf{X}_{n-1}-\mathbf{X}_{\t{target}}||}{U_{\t{swim}}} - \frac{||\mathbf{X}_n-\mathbf{X}_{\t{target}}||}{U_{\t{swim}}}\right]\\ +   \t{bonus}.
\end{split}
\end{equation}

\noindent
The first term penalizes duration of an episode to encourage fast navigation to the target. The second two terms give a reward when the swimmer is closer to the target than it was in the previous time step. The final term is a bonus equal to 200 seconds, or approximately 30 times the duration of a typical trajectory. The bonus is awarded if the swimmer successfully reaches the target. Swimmers that exit the simulation area or collide with the cylinder are treated as unsuccessful. The second two terms are scaled by 10 to be on the same order of magnitude as the first term, which we found significantly improved training speed and navigation success rates. We also investigated a non-linear reward function, in which the second two terms are the reciprocal of the distance to the target, however it exhibited lower performance. The RL algorithm seeks to maximize the total reward, which is the sum of the reward function across all $N$ time steps in an episode:

\begin{equation} \label{eq:rtotal}
    r_\t{total} = \sum_{n=1}^N r_n = -T_{\t{f}} + 10\frac{||\mathbf{X}_{\t{start}}-\mathbf{X}_{\t{target}}||}{U_\t{swim}} + \t{bonus}.
\end{equation}

\begin{figure}[!t]
\includegraphics[width=8.64cm]{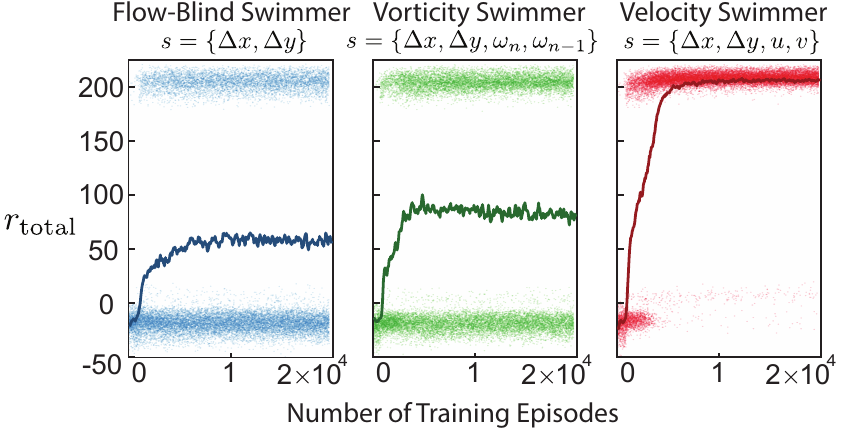}
\caption{Evolution of the cumulative reward during training for the three RL swimmers. The cumulative rewards for each episode are plotted as points, and a moving average with a window of 201 episodes is plotted with a solid line. Because the swimmer gains a bonus of 200 for reaching the target, successful episodes are clustered around a reward of 200 while unsuccessful episodes are clustered below zero.}
\label{fig:training}
\end{figure}

Assuming the swimmer reaches the target location, the only term in $r_\t{total}$ that depends on the swimmer's trajectory is $-T_{\t{f}}$. Therefore, maximizing the cumulative reward of a successful episode is equivalent to finding the minimum time path to the target. During training however, all terms in the reward contribute to finding policies that drive the swimmer to the target in the first place. The evolution of the reward function during training for each swimmer is shown in Figure \ref{fig:training}. All RL swimmers were trained for 20,000 episodes.

\begin{figure}[!ht]
\includegraphics[width=8.64cm]{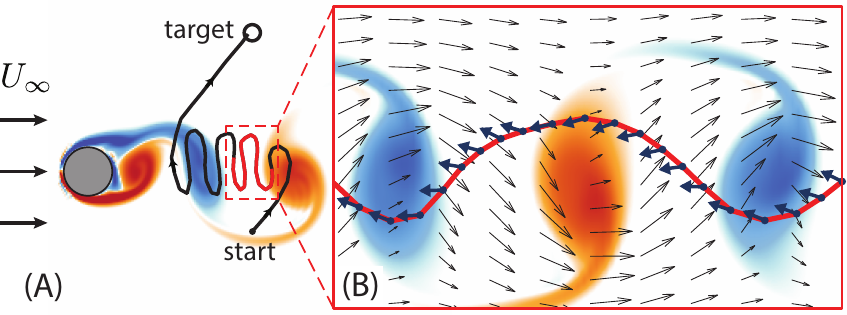}
\caption{(A) Example trajectory of the velocity RL swimmer, in which it successfully navigates from its starting location to the target. (B) Segment of this trajectory plotted in a wake-stationary frame of reference on top of the background flow field, which highlights the swimmer exploiting low-velocity regions in the cylinder wake to swim upstream. The swimming direction is plotted at each time step along the trajectory, revealing that this RL swimmer adjusts it swimming direction in response to the changing background flow, enabling time-efficient navigation.}
\label{fig:swimming_strategies}
\end{figure}

\begin{figure}[t]
\includegraphics[width=8.64cm]{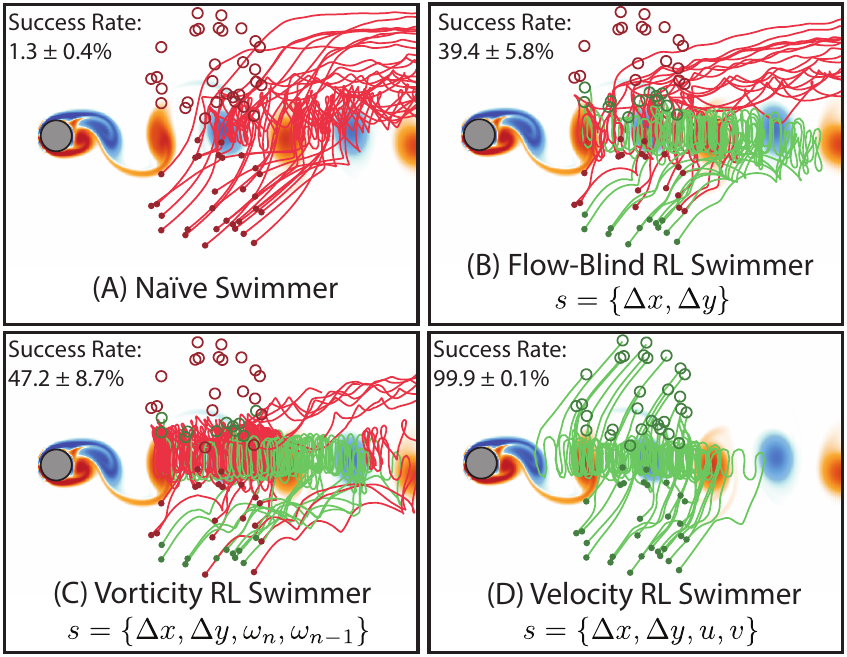}
\caption{Average success rate with 30 example trajectories for each swimmer type. Successful attempts to reach the target are green, while unsuccessful attempts are red. (A) Na\"{i}ve policy of swimming towards the target is rarely successful. (B) The flow-blind RL swimmer navigates more effectively than the na\"{i}ve swimmer. (C) The vorticity RL swimmer is more successful than the flow-blind swimmer, showing that sensing the local flow can improve RL-based navigation. (D) Surprisingly, the velocity RL swimmer nearly always reaches the target using only the local flow velocity. The stated success rates are averaged over 12,500 episodes and are shown with one standard deviation arising from the five times each swimmer was trained.}
\label{fig:success_rate}
\end{figure}

\emph{Success of RL Navigation}.\textemdash After training, Deep RL discovered effective policies for navigating through this unsteady flow. An example of a path discovered by the velocity RL swimmer is shown in Figure \ref{fig:swimming_strategies}. Because the swimming speed is less than the free-stream velocity, the swimmer must utilize the wake region where it can exploit slower background flow to swim upstream. Once sufficiently far upstream, the swimmer can then steer towards the target. The plot of the swimming direction inside the wake (Figure \ref{fig:success_rate}B) shows how the swimmer changes its swimming direction in response to the background flow, enabling it to maintain its position inside the wake region and target low-velocity regions.

However, the ability of Deep RL to discover these effective navigation strategies depends on the type of local flow information included in the swimmer state. To illustrate this point, example trajectories and the average success rates of the flow-blind, vorticity, and velocity RL swimmers are plotted in Figure \ref{fig:success_rate}, and are compared with a na\"{i}ve policy of simply swimming towards the target ($\theta_{\t{na}\text{\"i}\t{ve}} = \tan^{-1}\left({\Delta y / \Delta x}\right)$). 

A na\"{i}ve policy of swimming towards the target is highly ineffective. Swimmers employing this policy are swept away by the background flow, and reached the target only 1.2\% of the time on average. A reinforcement learning approach, even without access to flow information, is much more successful: the flow-blind swimmer reached the target locations nearly 40\% of the time. 

Giving the RL swimmers access to local flow information increases the success further: the vorticity RL swimmer averaged a 47.2\% success rate. Surprisingly however, the velocity swimmer has a near 100\% success rate, greatly outperforming the zebrafish-inspired vorticity approach. With the right local flow information, it appears that an RL approach can navigate nearly without fail through a complex, unsteady flow field. However, the question remains as to why some flow properties are more informative than others.

\begin{figure}[!ht]
\includegraphics[width=8.64cm]{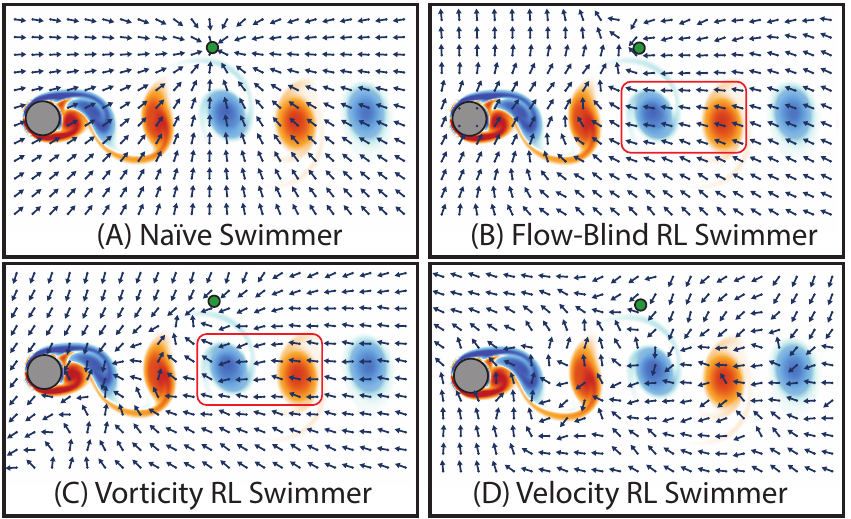}
\caption{Swimming direction policy plotted across the domain for a fixed target (green circle) at a given time instant. (A) The naïve swimmer swims towards the target. (B) The red outline highlights how the flow-blind swimmer navigates irrespective of the background flow, while the vorticity swimmer (C) adjusts its swimming direction modestly. (D) The velocity swimmer appears even more sensitive to the unsteady background flow.
}
\label{fig:policy_fields}
\end{figure}

To better understand the difference between RL swimmers with access to different flow properties, the swimming direction computed by each RL policy is plotted over a grid of locations in Figure \ref{fig:policy_fields}. The flow-blind swimmer does not react to changes in the background flow field, although it does appear to learn the effect of the mean background flow, possibly through correlation between the mean flow and the relative position of the swimmer in the domain. This provides it an advantage over the na\"{i}ve swimmer. The vorticity swimmer adjusts its swimming direction modestly in response to changes in the background flow, for example by swimming slightly upwards in counter-clockwise vortices and slightly downwards in clockwise vortices. The velocity swimmer appears most sensitive to the background flow, which may help it respond more effectively to changes in the background flow. 

Station-keeping inside the wake region may be important for navigating through this flow. In the upper right of the domain, the velocity swimmer learns to orient downwards and back to the wake region, while the other swimmers swim futilely towards the target. Because the vorticity depends on gradients in the background flow, that property cannot be used to respond to flow disturbances that are spatially uniform. These difference appear to explain many of the failed trajectories in Figure \ref{fig:success_rate}, in which the flow-blind and vorticity swimmers are swept up and to the right by the background flow.

It is worth noting that because the flow pushes the swimmers according to linear dynamics (Equation \ref{eq:dynamics}), the local velocity can exactly determine the swimmer's position at the next time step. This may explain the high navigation success of the velocity swimmer, as it has the potential to accurately predict its next location. To be sure, the Deep RL algorithm must still learn where the most advantageous next location ought to be, as the flow velocity at the next time step is still unknown.

While sensing of vorticity is insufficient to detect spatially uniform disturbances, it can be useful for distinguishing the vortical wake from the freestream flow. This can explain why the vorticity swimmer performs better than the flow-blind swimmer. A similar reasoning could apply to swimmers that sense other flow quantities such as pressure or shear. 

For real swimmers however, vorticity may play a larger role, for example by causing a swimmer to rotate in the flow \cite{colabrese_flow_2017} or by altering boundary layers and skin friction drag \cite{verma_efficient_2018}. Real robots would also be subject to additional sources of complexity not considered in this simplified simulation, which would make it more difficult to determine a swimmer's next position from local velocity measurements.

\newpage
\emph{Comparison with Optimal Control}.\textemdash In addition to reaching the destination successfully, it is desirable to navigate to the target while minimizing energy consumption or time spent traveling. Biferale et. al \cite{biferale_zermelos_2019} demonstrated that RL can approach the performance of time-optimal trajectories in steady flow for fixed start and target positions. Here, we find that this result also holds for the more challenging problem of navigating unsteady flow with variable start and target points.

As noted in Equation \ref{eq:rtotal}, maximizing $r_{\t{total}}$ is equivalent to minimizing the time spent traveling to the target ($T_f$), provided the swimmer successfully reaches the target. Therefore, we compare the velocity RL swimmer to the time-optimal swimmer derived from optimal control.

To find time-optimal paths through the flow, given knowledge of the full velocity field at all times, we constructed a path planner that finds locally optimal paths in two steps. First, a rapidly-exploring random tree algorithm (RRT) finds a set of control inputs that drive the swimmer from the starting location to the target location, typically non-optimally \cite{lavalle_randomized_2001}. Then we apply constrained gradient-descent optimization (i.e. the fmincon function in MATLAB) to minimize the time step (and therefore overall time $T_f$) of the trajectory while enforcing that the swimmer starts at the starting point (Equation \ref{eq:start}), obeys the dynamics at every time step in the trajectory (Equation \ref{eq:dynamics}), and reaches the target ($||\mathbf{X}_N-\mathbf{X}_{\t{target}}|| <= 0.1$). The trajectories produced by this method are local minima, so to approximate a globally optimal solution, we run the path planner 30 times and chose the fastest trajectory. Unlike with steady flow, Zermelo's classical solution for optimal navigation is not readily applicable for unsteady flow. Other algorithms could also be used to find optimal trajectories for unsteady flow given knowledge of the entire flow field \cite{lolla_time-optimal_2014,kularatne_going_2018}.

\begin{figure}[h!]
\includegraphics[width=8.64cm]{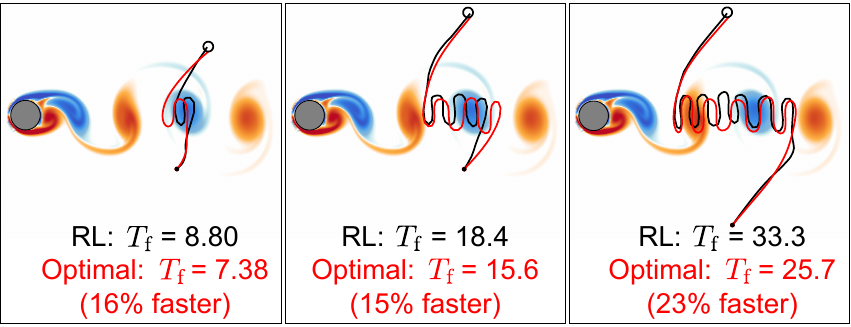}
\caption{Comparison between time-optimal trajectories (red) and RL trajectories (black) using an RL swimmer with state $s = \{\Delta x, \Delta y\,u,v\}$. Time to reach the target $T_{\t{f}}$ is made non-dimensional using the timescale $D/U_{\infty}$. }
\label{fig:optimal}
\end{figure}

A comparison between RL and time-optimal navigation for three sets of start and target points is shown in Figure \ref{fig:optimal}. These points were chosen to represent a range of short and long duration trajectories. Despite only having access to local information, the RL trajectories are nearly as fast and qualitatively similar to the optimal trajectories, which were generated with the advantage of having full global knowledge of the flow field. 

The surprisingly high performance of the RL approach compared to a global path planner suggests that deep neural networks can, to some extent, approximate how local flow at a particular time impacts navigation in the future. In other words, a successful RL swimmer must simultaneously navigate and identify the approximate current state of the environment. In comparison, the optimal control approach relies on knowledge of the environment in advance. There are limitations to the RL approach, however. For example, the optimal swimmer in the middle of Figure \ref{fig:optimal} enters the wake region at a different location than the RL swimmer to avoid a high velocity region, which the RL swimmer may not have been able to sense initially.

In addition to approaching the optimality of a global planner, RL navigation offers a robustness advantage. As noted in \cite{biferale_zermelos_2019}, RL can be robust to small changes in initial conditions. Here, we show that RL navigation can generalize to a large area of initial and target conditions as well as random starting times in the unsteady flow. RL navigation may also generalize to other flow fields to some extent \cite{colabrese_flow_2017}. In contrast, the optimal trajectories here are open loop: any disturbance or flow measurement inaccuracy would prevent the swimmer from successfully navigating the target. While robustness can be included with optimal control in other ways \cite{panda_comprehensive_2020}, responding to changes in the surrounding environment is the driving principle of this RL navigation policy. Indeed, the related algorithm of imitation learning has been applied to add robustness to existing path planners \cite{riviere_glas_2020}.

\newpage
\emph{Conclusion}.\textemdash We have shown in this Letter how Deep Reinforcement Learning can discover robust and time-efficient navigation policies which are improved by sensing local flow information. A bio-inspired approach of sensing the local vorticity provided a modest increase in navigation success over a position-only approach, but surprisingly the key to success was discovered to lie in sensing the velocity field, which more directly determined the future position of the swimmer. This suggests that RL coupled with an on-board velocity sensor may be an effective tool for robot navigation. Future investigation is warranted to examine the extent to which the success of the velocity approach extends to real-world scenarios, in which robots may face more complex, 3D fluid flows, and be subject to non-linear dynamics and sensor errors.

\bibliography{biblio.bib}

\end{document}